\documentclass[reprint,amssymb, amsmath, aip]{revtex4-1}
\usepackage{graphicx}
\usepackage{float}
\usepackage{color}
\usepackage{epsfig}
\usepackage{docs}%
\usepackage{bm}%
\usepackage[colorlinks=true,linkcolor=blue]{hyperref}%
\expandafter\ifx\csname package@font\endcsname\relax\else
 \expandafter\expandafter
 \expandafter\usepackage
 \expandafter\expandafter
 \expandafter{\csname package@font\endcsname}%
\fi
\begin{document}
\title{ Interaction and propagation characteristics of two counter and co-propagating Mach cones in a dusty plasma}
\author{P. Bandyopadhyay}%
\email{pintu31@gmail.com}
\author{R. Dey}
\author{A. Sen}
\affiliation{ Institute For Plasma Research, Bhat, Gandhinagar,Gujarat, India, 382428}%
\date{\today}
\begin{abstract}
We theoretically investigate the interaction and propagation characteristics of two co/counter propagating Mach cones triggered by two projectile particles moving with supersonic velocities in the same/opposite directions through a dusty plasma medium. The Mach cone solutions are obtained by solving a model set of fluid equations for a heavily charged dust fluid that includes the contributions of the projectile particles in the Poisson equation. The density profiles and velocity vector maps of the Mach wings show interesting structural changes when they interact with each other and form patterns similar to interference fringes.  Compared to the co-propagating Mach cones, the wings of counter propagating Mach cones produce a larger number of maxima and minima in the pattern resulting from their mutual interaction. In addition the time duration of the formation of two maxima or minima at a particular point decreases due to the interactions of Mach cones. Another notable feature is that the spacing between adjacent maxima increases while the fringe angle decreases with the increase of relative velocity of the counter propagating projectile particles. 
\end{abstract}
\maketitle
\section{Introduction} 
Wakes, consisting of expanding $V$ shaped wave structures, are a common occurrence behind objects moving in a fluid such as a boat on a lake surface. In this case they consist of linear surface water waves that are excited by the perturbation caused by the moving boat and they move away in the direction perpendicular to the motion of the boat at their characteristic phase velocity. A similar phenomenon occurs when an airplane moves through the atmosphere creating a three dimensional structure of wake fields behind it. The two dimensional projection of such a trailing structure again has a characteristic $V$ shape. When the speed of the moving object exceeds the sound speed of the medium the outer boundary of the cone shaped structure is defined by a propagating shock wave excited by the moving object and the structure is then termed as a Mach cone. Wake fields and Mach cones have been widely studied in fluid and gas dynamics \cite{Heil,Funfschilling,Joseph,Bond} as well as in plasmas \cite{Havnes1,Havnes2,Dubin,Ma,Zhdanov1,Mamun,Hou1,Jiang1,Hou2,Bose,Zhdanov2,Pintu1,Pintu2,Thomas2,Samsonov1,Samsonov2,Melzer,Nosenko1,Nosenko2,Jiang2,Mierk} for their intrinsic properties as well as their practical applications. In plasmas, laser pulse induced wake-fields have generated a great deal of interest due to their potential as a high energy particle accelerator \cite{Melzer,Nosenko1,Nosenko2}. More recently Mach cones excited by a supersonically moving charged particle in a dusty plasma medium have received some attention \cite{Samsonov1,Samsonov2,Jiang2,Mierk}.  
A dusty (or complex) plasma \cite {Ikezi,Thomas,Hayashi,Chu} consists of micron sized dust grains that are embedded in an electron-ion plasma. The massive dust particles (that can range in size from a few nanometers to micrometers) get negatively charged by acquiring more electrons than ions from the background and their dynamics leads to new low frequency collective modes in the medium. Since dust particles are of a macroscopic size they are visible to the eye and their dynamics can be easily tracked using visual techniques. This characteristic feature of a dusty plasma has motivated and facilitated  a large amount of theoretical and laboratory experimental work on linear and nonlinear wave propagation studies including the study of wakes and Mach cones \cite{Havnes1,Havnes2,Dubin,Ma, Zhdanov1, Mamun, Hou1, Jiang1,Hou2,Bose, Zhdanov2, Pintu1, Pintu2,Thomas2,Samsonov1,Samsonov2,Melzer,Nosenko1,Nosenko2}. 
As is well known, in a Mach cone the angle formed by the two shock fronts ($\theta$) is related to the Mach number $M$ by the relation $sin(\theta) = 1/M$. Measurement of the cone angle thus provides a convenient means of assessing the speed of the travelling particle and the utility of this simple diagnostic tool in the context of dusty plasmas has been pointed out by Havnes et al \cite{Havnes2}. An interesting extrapolation from laboratory observations of Mach cones in dusty plasmas has been made by  Brattli {\it et al}  \cite{Brattli} who have predicted the existence of Mach cone structures in Saturn's ring region due to fast moving large sized charged boulders moving midst a dusty plasma composed of finer sized particles and an electron-ion plasma. Detection of such a Mach cone pattern from space platforms like satellites can provide valuable information about the speed and location of the revolving boulders. However, in reality it is likely that the Mach cone generated by a given boulder would interact and interfere with similar cones generated by neighbouring boulders resulting in a more complex wave pattern. One commonly sees emergence of such patterns in water waves when two boats travel close to each other or pass each other from opposite directions. The emergent interference patterns have characteristic spatio-temporal properties including pockets of intense wave amplitudes. It is of interest therefore to investigate the nature of such dynamical patterns in a dusty plasma when two Mach cones are simultaneously generated by charged objects travelling in the same or opposite directions to each other. We have carried out such an investigation in this paper by a theoretical analysis of a model system consisting of a dusty plasma fluid in which two charged sources travel at supersonic speeds (with reference to the dust acoustic speed) over a range of separation distances and differing directions. 

The paper is organized as follows. In Sec.~\ref{sec:theory}, we describe the governing equations associated with the model used in our study and present analytic solutions representing the linear perturbed density and velocity vector mapping of two counter/co-propagating Mach cones. Numerical evaluations of these solutions provide the spatio-temporal dynamics of the cones and these are displayed and discussed in Sec.~\ref{sec:results}. A summary of our results and some brief concluding remarks are given in Sec.~\ref{sec:conclusion}.     
\section{Theoretical model}\label{sec:theory}
A standard approach for studying low frequency phenomena in a dusty plasma, in the regime where dust dynamics is important, is to model the dynamics of the charged massive dust component by the following set of fluid equations consisting of the continuity, the momentum and the Poisson equation (Eq.~\ref{eqn:con}--\ref{eqn:pos}):
\begin{eqnarray} 
&&\frac{\partial n_d(\mathbf{r},t)}{\partial t}+\nabla.[n_d(\mathbf{r},t)\mathbf{v_d}(\mathbf{r},t)]=0,\label{eqn:con}\\ 
&&\frac{\partial \mathbf{v_d}(\mathbf{r},t)}{\partial t}+\mathbf{v_d}\nabla. \mathbf{v_d}(\mathbf{r},t)=\frac{Z_{d}e}{m_d}\nabla \phi(\mathbf{r},t)+F_{EP}, \label{eqn:mom}\\ 
&&\nabla^2\phi(\mathbf{r},t)=\frac{e}{\epsilon_0}[Z_dn_d(\mathbf{r},t) + Z_t\delta(\mathbf{r}+\mathbf{r_{01}}-\mathbf{u_{1}}t) \nonumber\\ 
&& \hspace{0.5in} + Z_t\delta(\mathbf{r}+\mathbf{r_{02}}-\mathbf{u_{2}}t)+n_e(\mathbf{r},t) - n_i(\mathbf{r},t)].
\label{eqn:pos}
\end{eqnarray}
\noindent
In the above equations, $n_d(\mathbf{r},t)$, $\mathbf{v_d}(\mathbf{r},t)$ and $\phi(\mathbf{r},t)$ represent the instantaneous number density, velocity vector and the electrostatic potential of the dust fluid, respectively in a 2D Cartesian coordinate system with $\mathbf{r} (x,y,z=0)$. $m_d$ denotes the mass of dust particles and $Z_d$($=\frac{Q_d}{e}$) is the charge number acquired by a single dust where $e$ is the elementary charge of an electron. In our calculations systematic or stochastic dust charge variations are neglected. The electrostatic ($\mathbf{F}_E=Z_{d}e\nabla \phi(\mathbf{r},t)$) and Epstein drag ({$\mathbf{F}_{EP}$}) forces which are acting on the dust particles are taken into account in the momentum equation (Eq.~(\ref{eqn:mom})). The Epstein drag force arises due to collision between the background neutral gas molecules and dust particles, and can be expressed as, \cite{Epstein} 
\begin{eqnarray}
\mathbf{F}_{EP} &=& -\gamma_{EP} \mathbf{v_{d}}(\mathbf{r},t) \nonumber \\& =& -\delta_{EP}\frac{4\pi}{3}N_n\frac{m_n}{m_d}v_nr_d^2\mathbf{v_{d}}(\mathbf{r},t).
\end{eqnarray}
\noindent
Where $\gamma_{EP}$ is the Epstein drag coefficient. $N_n$, $m_n$ and $v_n$ are the density, mass and thermal velocity of background neutrals, respectively. $r_d$ is the radius of the dust particle.\par
\begin{figure}[!]
\includegraphics[width=0.49\textwidth]{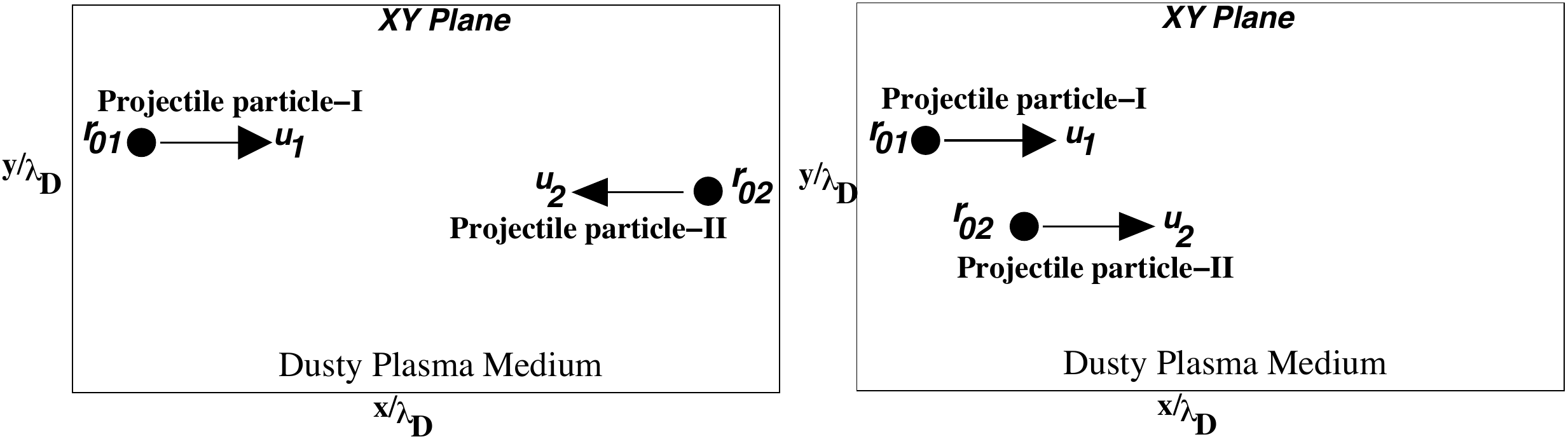}
\caption{Initial locations and velocities of two projectile particles in case of (a) counter-propagating and (b) co-propagating Mach cones}
\label{figure0}
\end{figure}
It may be noted that the contributions of the two projectile particles arise through their individual charge density perturbations and these have been incorporated on the right hand side of the Poisson's equation (Eq.~(\ref{eqn:pos})) in terms of delta functions. The arguments of the delta functions take into account the magnitudes of their different velocities $\mathbf{u_{1}}$ and $\mathbf{u_{2}}$ as well as their directions of propagation. They are both assumed to carry the same charge $Z_t e$. $\mathbf{r_{01}}$ and $\mathbf{r_{02}}$ are the initial positions of the projectile dust particles as shown in Fig.~\ref{figure0}. The direction of the projectile particle can be changed by changing the sign of $\mathbf{u_{1}}$ and $\mathbf{u_{2}}$. To excite two counter-propagating Mach cones, we assume the velocity of the first projectile particle is positive (moves from left to right) whereas the second particles move with a negative velocity (right to left). In the case of co-propagating Mach cones, both the projectile particles are always chosen to move from left to right with positive velocities of different values. The trailing particle is chosen with a faster velocity than the leading particle so that in the course of time the Mach cone of the trailing particle can catch up with the Mach cone of the leading particle and interact with it.  \par
Since the electrons and ions are both lighter species compared to the dust fluid, their dynamics is over a much faster time scale and they can be modeled by Boltzmannian distributions with characteristic temperatures $T_e$ and $T_i$, respectively. The densities of the electrons ($n_e$) and ions ($n_i$) can be then given by,
\begin{eqnarray}
n_e&=&n_{e0}\text{exp}(e\phi(\mathbf{r},t)/k_BT_e),\\
n_i&=&n_{i0}\text{exp}(-e\phi(\mathbf{r},t)/k_BT_i).
\end{eqnarray}
\noindent
where n$_{e0}$ (n$_{i0}$) is the equilibrium density of electrons (ions) and $k_B$ is the Boltzmann constant. The equilibrium electron density $n_{e0}$ and ion density $n_{i0}$ are related to the equilibrium dust density $n_{d0}$ and the dust charge number $Z_d$ by the charge neutrality condition,
\begin{eqnarray}  
n_{i0}=n_{e0}+n_{d0}Z_d.
\label{eqn:quasi}
\end{eqnarray}
For a perturbed situation, assuming a first order approximation, the dynamical variables $n_d(\mathbf{r},t)$, $\mathbf{v_d}(\mathbf{r},t)$, $\phi(\mathbf{r},t)$, $n_e(\mathbf{r},t)$ and $n_i(\mathbf{r},t)$ about the unperturbed states are given by,
\begin{eqnarray}
\nonumber
&n_d(\mathbf{r},t)&=n_{d0}+n_{d1}(\mathbf{r},t),\label{eqn:nd}\\ \nonumber
&\mathbf{v_d}(\mathbf{r},t)&= \mathbf{v_{d1}}(\mathbf{r},t),\label{eqn:vd}\\
&\phi(\mathbf{r},t)&=\phi_1(\mathbf{r},t) ,\label{eqn:phi}\\\nonumber
&n_e(\mathbf{r},t)&=n_{e0}+n_{e0}\left(\frac{e}{k_BT_e}\right)\phi_1(\mathbf{r},t),\label{eqn:ne}\\\nonumber
&n_i(\mathbf{r},t)&=n_{i0}-n_{i0}\left(\frac{e}{k_BT_i}\right)\phi_1(\mathbf{r},t).\label{eqn:ni}
\end{eqnarray}
Using Eqs. (\ref{eqn:quasi}) and (\ref{eqn:phi}) in Eqs. (\ref{eqn:con}), (\ref{eqn:mom}) and (\ref{eqn:pos}) we obtain the following set of linearized equations,
\begin{eqnarray}  
&&\frac{\partial n_{d1}({\bf{r}},t)}{\partial t} + n_{d0}\nabla {.{\bf{v}}_{d1}({\bf{r}},t)}=0,\\
&&\frac{\partial {\bf{v}}_{d1}({\bf{r}},t)}{\partial t}=\frac{Z_de}{m_d}\nabla \phi_1({\bf{r}},t)-\gamma_{EP}{\bf{v}}_{d1}({\bf{r}},t)\,\\
&&\nabla^2 \phi_1({\bf{r}},t)=4\pi e[Z_dn_{d1}({\bf{r}},t)+Z_t\delta({\bf{r}+\bf{r_{01}}}-{\bf{u_{1}}}t)\nonumber\\
&&\hspace{0.7in}+Z_t\delta({\bf{r}+\bf{r_{02}}}-{\bf{u_{2}}}t)]+\lambda_{D}^{-2}\phi_1({\bf{r}},t).
\end{eqnarray}
\noindent
Where 
\begin{eqnarray}  
\lambda_{D}=\left[\frac{\epsilon_{0} k_{B}}{e^{2}}\left(\frac{T_{e} T_{i}}{T_{i} n_{e0}+T_{e} n_{i0}}\right)\right]^{1/2} 
\label{eqn:lam1}
\end{eqnarray}
is the  Debye length in a dusty plasma.
\par
We next take a Fourier transform in space and time of the above equations to obtain a set of algebraic equations, where the transform of quantity $A(\mathbf{r},t)$ is defined as,
\begin{eqnarray}  
A({{\bf{r}},t)}=\int{\frac{d{\bf{k}}d\omega}{(2\pi)^4}A({\bf{k}},\omega)e^{i{\bf{k} \cdot r}-i\omega t}}.
\label{eqn:Fourier}
\end{eqnarray}
The algebraic equations obtained from the Fourier transformed form of Eq.~(\ref{eqn:Fourier}) can be easily solved for $\tilde{n}_{d1}(\mathbf{k},\omega)$, $\mathbf{\tilde{v}_{d1}}(\mathbf{k},\omega)$ and $\tilde{\phi}_{1}(\mathbf{k},\omega)$. Taking the inverse Fourier transform, we finally obtain,
\begin{eqnarray}  
&n_{d1}&(\mathbf{{r_{1}}},\mathbf{{r_{2}}},t)=\frac{\beta}{(2\pi)^4}\int{d\mathbf{k}d\omega \frac{[1-\epsilon(k,\omega)]}{\epsilon(k,\omega)}} \\ \nonumber  &&\hspace{-0.1in}\times \left[e^{i({\mathbf{{k} \cdot {r_{1}}}}-\omega t)} \delta(\omega-\mathbf{k \cdot u_{1}})+ e^{i({\mathbf{{k} \cdot {r_{2}}}}-\omega t)}\delta(\omega-\mathbf{k \cdot u_{2}})\right ], \label{eqn:density}
\end{eqnarray}
\begin{eqnarray}
&\mathbf{v}_{d1}&(\mathbf{{r_{1}}},\mathbf{{r_{2}}},t)=\frac{\beta}{(2\pi)^4n_{d0}}\int{d{\mathbf{k}}d\omega \frac{\omega}{k^2}\frac{[1-\epsilon(k,\omega)]}{\epsilon(k,\omega)}} \\ \nonumber &&\hspace{-0.1in} \times \left[e^{i({\mathbf{{k} \cdot {r_{1}}}}-\omega t)} \delta(\omega-\mathbf{k \cdot u_{1}})+e^{i({\mathbf{{k}.{r_{2}}}}-\omega t)} \delta(\omega-\mathbf{k \cdot u_{2}})\right]\mathbf{k},\label{eqn:velocity}
\end{eqnarray}
and
\begin{eqnarray}  
&\phi_1&(\mathbf{{r_{1}}},\mathbf{{r_{2}}},t)=\frac{(-4 \pi e Z_t {\lambda_{D}}^2)}{(2\pi)^4}\int{{d{\mathbf{k}}d\omega} \frac{1}{(1+k^2 {\lambda_D}^2)}\frac{1}{\epsilon(k,\omega)}} \nonumber\\& \times& \left[e^{i({\mathbf{{k} \cdot {r_{1}}}}-\omega t)} \delta(\omega-\mathbf{k \cdot u_{1}})+ e^{i({\mathbf{{k} \cdot {r_{2}}}}-\omega t)}\delta(\omega-\mathbf{k \cdot u_{2}}) \right]. \label{eqn:potential}
\end{eqnarray}
\noindent
where $\mathbf{{r_1}}=\mathbf{r}+\mathbf{r_{01}}$, $\mathbf{{r_2}}=\mathbf{r}+\mathbf{r_{02}}$ and the dielectric function $\epsilon(k,\omega)$ is given by,
\begin{eqnarray}  
\epsilon(k,\omega)=1-\frac{\omega_{pd}^2}{\omega\left[(\omega+i\gamma_{EP})\right]}\left(\frac{k^2\lambda_D^2}{k^2\lambda_D^2+1}\right) \label{eqn:die}
\end{eqnarray}
\noindent
with the dust plasma frequency $\omega_{pd}=(\frac{n_{d0}Z_d^2e^2}{\epsilon_0m_d})^{1/2}$ and the ratio of charges $\beta$= Z$_t$/Z$_d$. In the above, we have normalized all space coordinates by $\lambda_D$, time by the inverse of $\omega_{pd}$ and velocity by the dust acoustic speed $C_s=\lambda_D\omega_{pd}$.\par 
\begin{figure*}[!]
\includegraphics[width=0.90\textwidth]{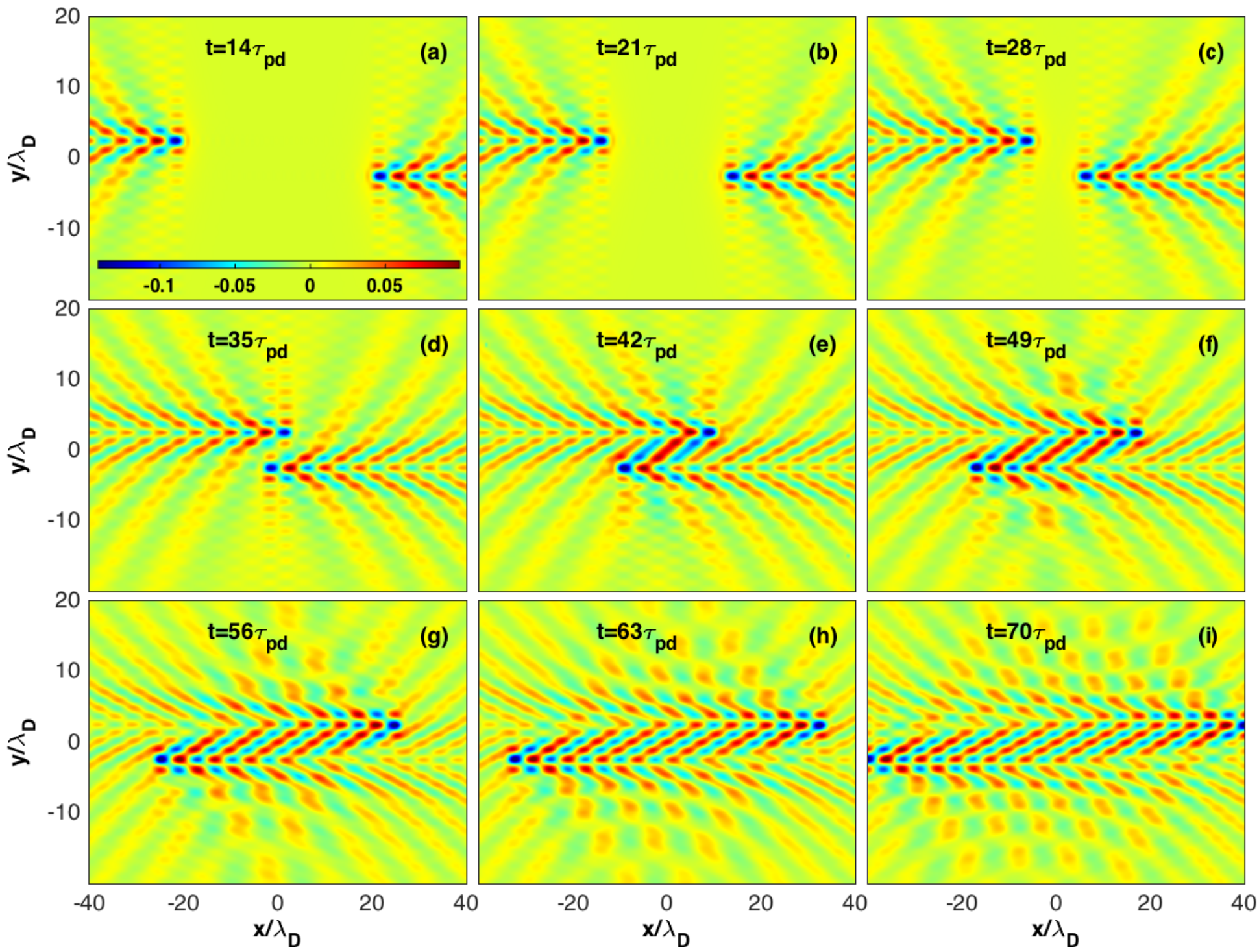}
\caption{ (Color online). Time evolution of the density ($n_{d1}$) profile for $\gamma_0$=0.01 and M$_{1}$=-1.1 and M$_{2}$=1.1.  The profiles before the interaction are plotted in (a--c) whereas (f--i) depicts the profiles after the interaction. (d) shows the density profiles when the wings of the Mach cones start to interact.}
\label{figure1}
\end{figure*}
To calculate the perturbed particle density, velocity vectors and potential, we solve these expressions (Eq. (\ref{eqn:density})--(\ref{eqn:potential})) numerically for plasma and dusty plasma parameters that are representative of typical laboratory dusty plasma experiments \cite{Thomas3}. The plasma density ($n_i$), the electron ($T_e$) and ion ($T_i$) temperatures are chosen as 10$^{14}$/m$^{3}$, 3 eV and 0.1 eV respectively, whereas, the dusty plasma parameters e.g., the charge (Q$_t$=Z$_t$e) acquired by a projectile particle, average dust charge (Q$_d$=Z$_d$e), dust diameter ($=2r_d$), inter-particle distance (d) are chosen as 10$^{5}$e, 40007e, 9.0$\mu$m and 230$\mu$m, respectively. Furthermore the velocities of the projectile particles moving in the opposite directions to generate the counter-propagating Mach cones are always chosen to have the same magnitudes. For co-propagating Mach cones the projectiles are given different values. In this study, the projectile particles are initially placed at locations $(x/\lambda_D=-35$, $y/\lambda_D=-2.5)$ and $(x/\lambda_D=35$, $y/\lambda_D=2.5)$ for the case of counter-propagating Mach cones and at $(x/\lambda_D=-30$, $y/\lambda_D=-2.5)$ and $(x/\lambda_D=-20$, $y/\lambda_D=2.5)$ for the case of co-propagating Mach cones. 
\begin{figure*}[!]
\includegraphics[width=0.80\textwidth]{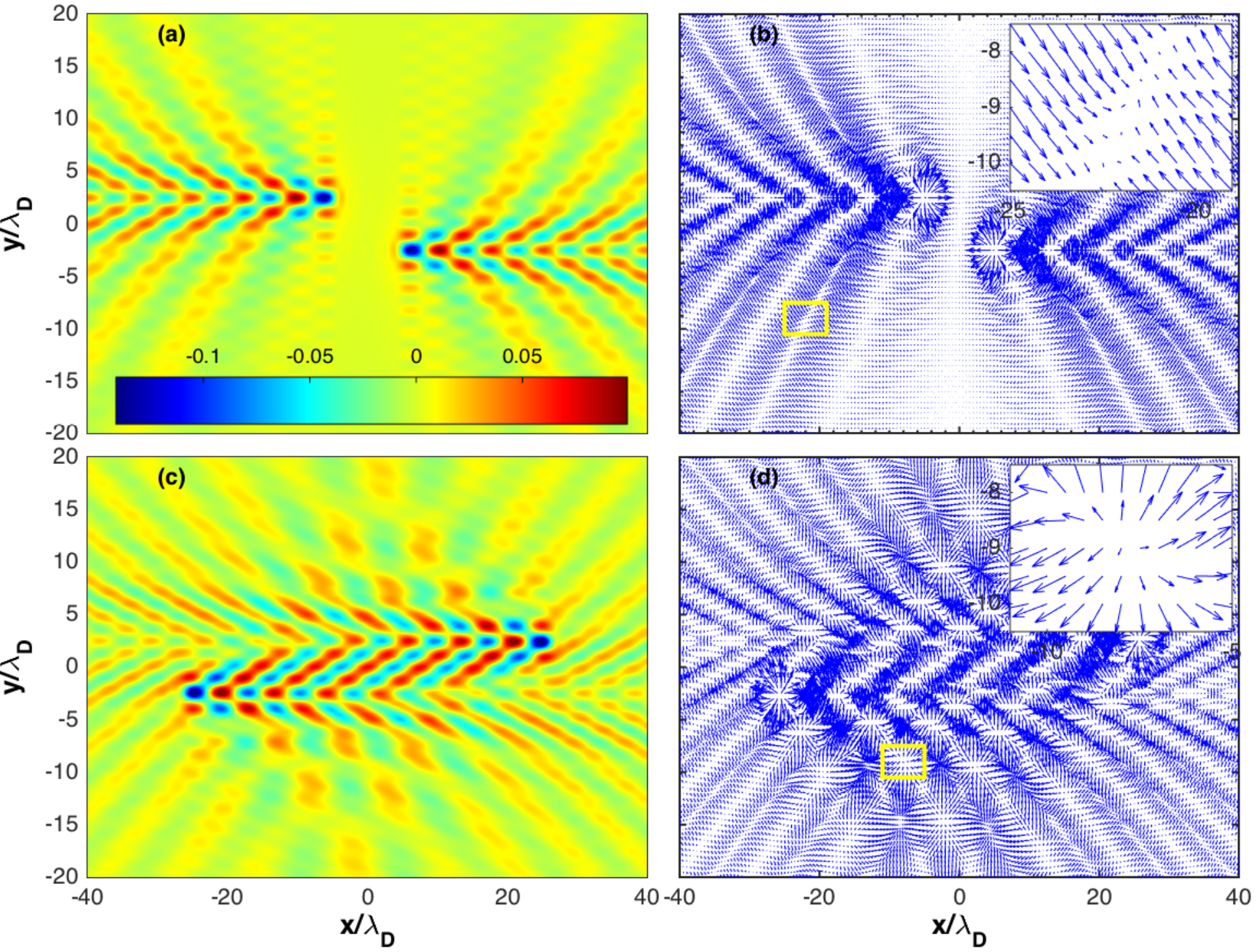}
\caption{(Color online). Density ($n_{d1}$) profiles and velocity vector ($v_{d1}$) maps before (a) \& b) and after (c) \& d) the interaction of the wings of two counter propagating Mach cones. This profiles are given for $\gamma_{0}=0.01$, M$_1=-1.1$ and M$_2=1.1$.}
\label{figure2}
\end{figure*}
\section{Results and Discussion}
\label{sec:results}
Fig.~\ref{figure1} shows the time evolution of density fluctuations (obtained from Eq.~14) of the wakes created by two projectile particles moving with the same velocity but in opposite directions (M$_1=-1.1$ and M$_2=1.1$, where M$_1$ and M$_2$ are the Mach numbers of projectile particles defined as the ratio of velocity of projectile particle to the dust acoustic velocity). The dust neutral damping is taken to be very weak ($\gamma_{0}=0.01$) in this case. The projectile particles are initially ($t=0$) placed at positions $(x/\lambda_D=35$, $y/\lambda_D=-2.5)$ and $(x/\lambda_D=-35$, $y/\lambda_D=2.5)$. As time progresses the travelling projectiles approach each other till at $t=35\tau_d$ (Fig.~\ref{figure1}(d)) they are nearly at the same position in $x/\lambda_D$. Where $\tau_d(=1/\omega_{pd})$, is the time taken by the dust particle for a complete oscillation. Beyond this time, as the projectiles progress further, the wings of their trailing Mach cones start to mutually interact with each other creating an interesting interference like pattern of fringes in the space between the trajectory of the two projectiles. This regular pattern of density maxima (red fringes) and minima (blue fringes) continues to form for a considerable amount of time (see Fig.~\ref{figure1}(f-i)) covering a large region in $x/\lambda_D$-space till the projectiles are far apart and the interaction between their Mach cones gets considerably weakened.  \par
\begin{figure*}[!]
\includegraphics[width=0.8\textwidth]{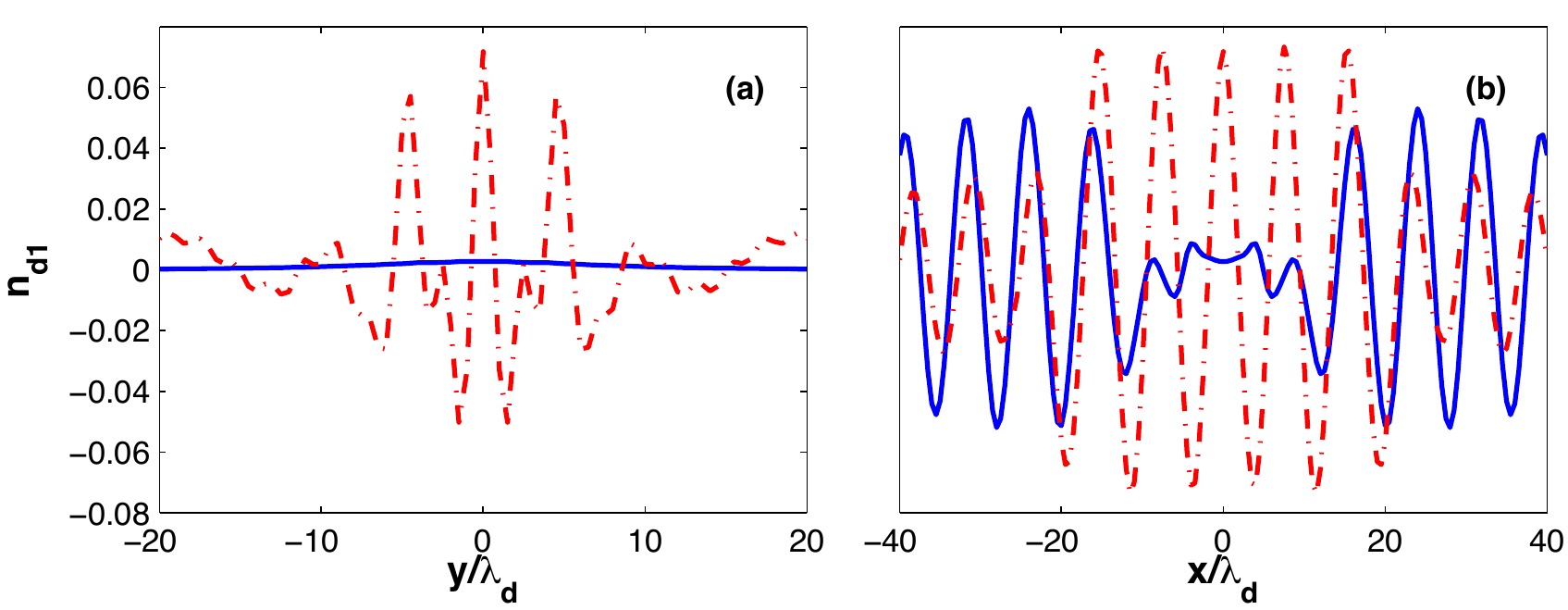}
\caption{(Color online). Variation of density along (a) y direction (perpendicular to which projectile particle moves) for a particular values of $x/\lambda_{D}$ (=0.0) before (solid line) and after the interaction (dashed line) and (b) x direction (along which projectile particle moves) for a particular values of $y/\lambda_{D}$ (=-10.0) before (solid line) and after the interaction (dash-dotted line). The normalised dust neutral collision frequency and Mach numbers for this plot are chosen, $\gamma_0=0.01$ and $M_1=-1.1$ and $M_2=1.1$, respectively.}
\label{figure3}
\end{figure*}
For a more detailed examination of the dynamics of interaction Fig.~\ref{figure2} depicts the perturbed density profiles ($n_{d1}$) (Fig.~\ref{figure2} (a) and (c)) and  velocity ($v_{d1}$) vector maps (Fig.~\ref{figure2} (b) and (d)) of the above described two counter propagating lateral wake structures. These snapshots are for two different times. In Fig.~\ref{figure2}(a) and (b), the wakes are about to interact with each other at a time $t=28\tau_d$, whereas in Fig.~\ref{figure2}(c) and (d), the wakes have interacted and propagated a distance of $x/\lambda_D \sim 70$ at time, $t=56\tau_d$. It should be mentioned that the structural properties of these wakes depend upon the wave dispersion properties of the fluid medium and on how strongly the projectile particles interact with the wave. Earlier observations \cite{Jiang1, Pintu1} also suggest that these V-shaped lateral wakes found behind each projectile (before their mutual interaction) are a consequence of the superposition of longitudinal dust acoustic waves and the multiple structures arise because of the strong dissipative nature of the waves. A significant change in the wing structures of the counter propagating Mach cones is observed in the post interaction phase. Maxima (red) and minima (blue) line fringes are seen near the source points in the perturbed density profile along the direction of the propagation of the Mach cone. In addition to these line fringes spots of maxima (red) and minima (blue) are also observed away from the source point in the y-direction. These maxima and minima spots indicate that the lateral wakes lose their identity beyond a certain distance in the post interaction phase. Similar interesting changes are also found in the velocity vector map as shown in Fig.~\ref{figure2}(b) and (d) and in the zoomed views (see in the insets) of the same regions for pre- and post-collisional cases. Before the interaction the velocity vectors show dark interference fringes whereas during the interaction they become point dark areas.\par %
For a quantitative measure of the interaction pattern, we have plotted the density profiles of two counter propagating Mach cones before (by solid lines) and after (dash-dotted lines) the interaction in Fig.~\ref{figure3}(a) and Fig.~\ref{figure3}(b). Fig.~\ref{figure3}(a) is the variation of density along the $y$ axis at a particular point $x/\lambda_D=0$ whereas, Fig.~\ref{figure3}(b) is the same along the $x/\lambda_D$ axis when the observation point is kept at $y/\lambda_D=-10$.  As seen in Fig.~\ref{figure2}(a), the Mach cones do not interact with each other and as a result the density fluctuations along the $y$-axis for a particular $x/\lambda_D=0$ are negligible which is shown in Fig.~\ref{figure3}(a) by a solid line. The fluctuations in density become prominent and oscillatory in nature when they interact and cross each other. Fig.~\ref{figure3}(b) shows the density profile variations along the x axis for a particular value of $y/\lambda_D$ ($=-10$) before (solid line) and during (dash-dotted line) the interaction phase. The oscillations in the solid line in the regions between $-40<x/\lambda_D<10$ and $10<x/\lambda_D<40$ capture amplitude variations of the individual wakes of the two projectiles as they approach each other. The nearly flat nature of the curve in the region $-4<x/\lambda_D<4$ indicates that the wakes have not yet reached this region and hence have not started to interact with each other. The dash-dotted curve shows high amplitude multiple peaks in this region as a result of the interaction between the wakes.  \par
\begin{figure*}[!]
\includegraphics[width=0.6\textwidth]{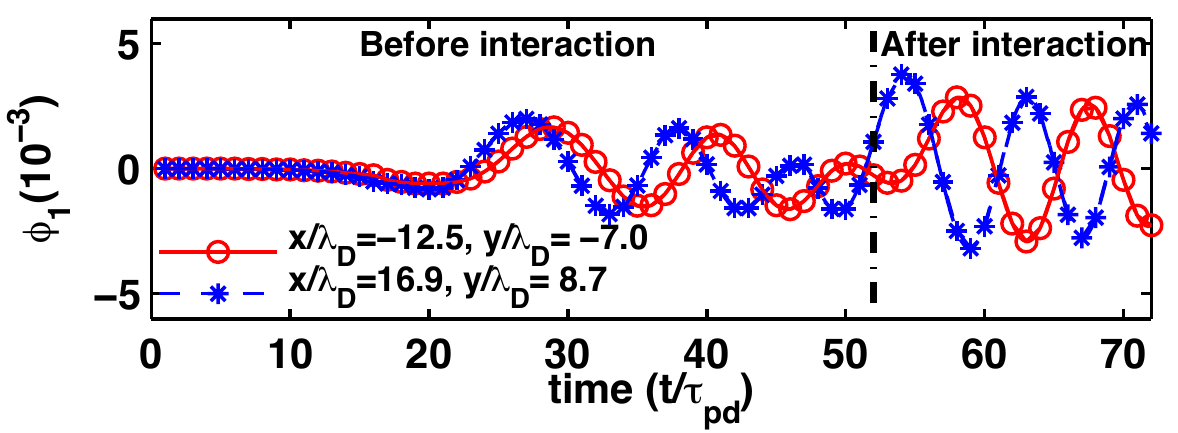}
\caption{(Color online) Variation of fluctuated potentials at two particular points as time evolves for $\gamma_0$=0.01 and M$_{1}$=-1.1 and M$_{2}$=1.1.  The solid line with open circles represents the potential for $x/\lambda_D=-12.5$, $y/\lambda_D=-7.0$ whereas the dashed line with stars represents for $x/\lambda_D=16.9$, $y/\lambda_D=8.7$. }
\label{figure3a}
\end{figure*}
\begin{figure*}[!]
\includegraphics[width=0.7\textwidth]{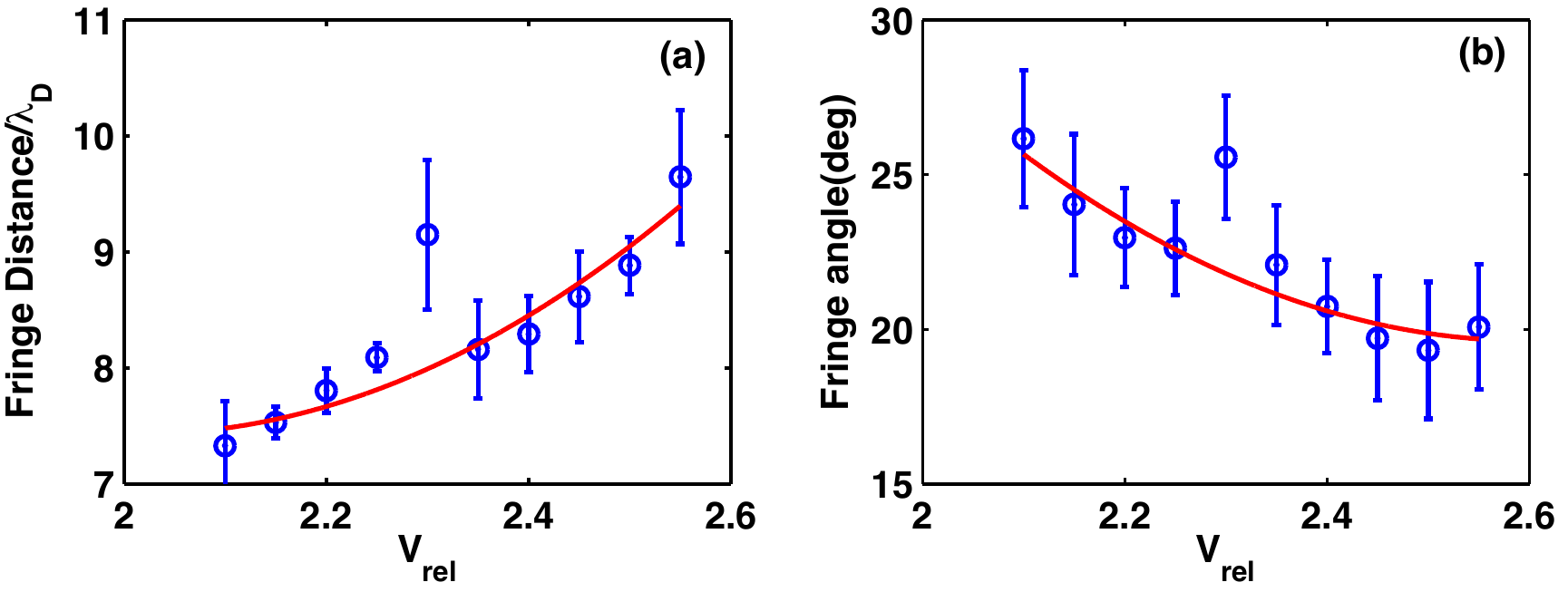}
\caption{(Color online). Variation of (a) fringe distance and the (b) fringe angle with the relative velocities of the projectile particles when they travel in opposite direction. The distance of the interference pattern and the fringe angles are calculated in each case at the time $t=64\tau_d$. The normalized dust neutral collision frequency is chosen as $\gamma_0$=0.01.}
\label{figure4}
\end{figure*}
\begin{figure*}[ht]
\includegraphics[width=0.9\textwidth]{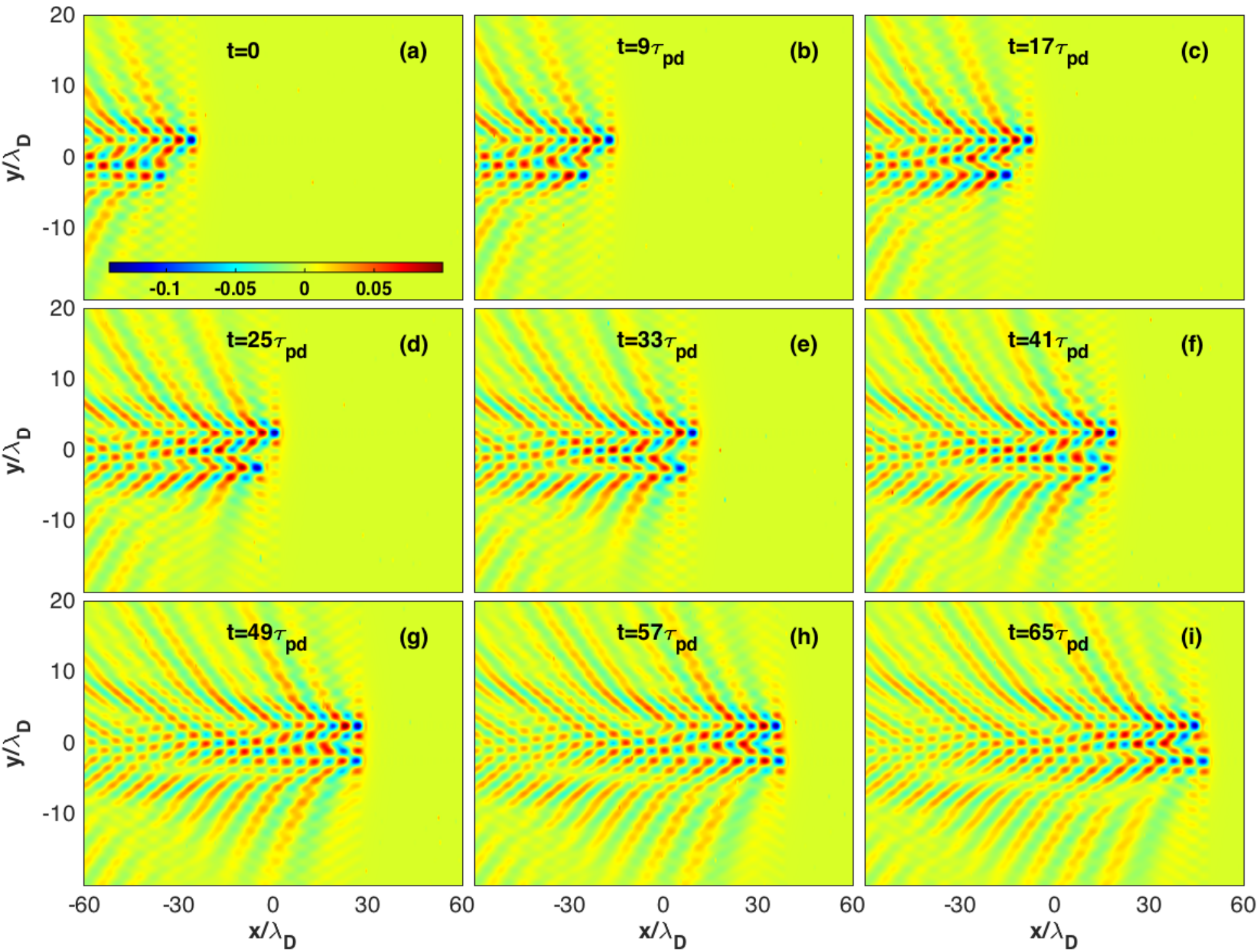}
\caption{(Color online). Time evolution of the density profiles of two co-propagating Mach cones for $\gamma_0$=0.01. In this case, the trailer projectile particle is moving with higher velocity ($M=1.3$), whereas the leading projectile particle is  propagates with smaller velocity ($M=1.1$). Fig.~(a--g) show the resultant potential profiles of two co-propagating  Mach cones when the faster projectile particle travels behind the slower projectile particle, whereas (h--i) depict the profiles after the faster projectile particle overtakes the slower one.}
\label{figure5}
\end{figure*}
To study the temporal evolution of the potential profiles, we have chosen two different spatial locations at $P_1$ (-12.5, -7.0) and $P_2$  (16.9, 8.7) on the x-y plane and plotted the values of the perturbed potential as time progresses. Fig.~\ref{figure3a} shows the variation of this potential with time ($t$) at these two points for $\gamma_0$=0.01 and M$_{1}$=-1.1 and M$_{2}$=1.1. The solid line with open circles represents the potential at  $P_1$ whereas the dashed line with stars is for $P_2$. This figure shows that the strength of the potential changes at a particular spatial point from maximum to minimum value as time evolves. The maxima and minima at the left of the dash-dotted line (placed at $t=52\tau_d$) is due to the propagation of individual Mach cone whereas at the right, it is due to the interaction of the wings of two Mach cones. It is clearly seen in the figure that the strength of potential increases due to the strong interactions of the wings of the Mach cones at the time of the formation of constructive interference patterns.  The time duration of the formation of two maxima at a particular point decreases from $\sim 12\tau_d$ to $\sim 10\tau_d$ when the Mach cones interact with each other. \par
Fig.~\ref{figure4}(a) shows the variation of distance of two constructive interference fringes produced in the resultant density profile with the relative velocity of the projectile particles when they travel in opposite directions. In this plot, both the projectile particles are chosen to travel with different velocities over the range M$_{1}$=1.05 to 1.25 in opposite directions, hence the relative velocity ($v_{rel}$) changes from 2.1 to 2.55. The distance between two neighbouring fringes in the resultant interference patterns is measured for all the cases at time $t=64\tau_d$ along the line $y/\lambda_D=0$. The normalized dust neutral collision frequency is chosen as $\gamma_0$=0.01. It is clearly seen from the figure that the distance between two neighbouring fringes increases with the increase of relative velocities of the projectile particles.  Fig.~\ref{figure4}(b)  displays the variation of the inclination of the fringe to the direction of propagation of the projectile (i.e. the angle between the interference pattern and the line $y/\lambda_D=0$) with the relative velocity of the projectile particles for the same condition of Fig.~\ref{figure4}(a). It clearly shows that the fringe angle decreases with the increase of the relative velocity of the projectile particle. These characteristic dependencies provide valuable insights into the dynamics of the interaction patterns and may prove  helpful as diagnostic tools for inferring the velocities of the source particles from a measurement of the fringe patterns. It is worth mentioning that the inter fringe distance and the fringe angle do not change significantly with the dust-neutral collision frequency (not shown in the figure).  \par
Fig.~\ref{figure5} shows the time evolution of the resultant density profiles of the wakes created by two projectile particles which are moving with different velocities in the same direction (M$_1$=1.1 and M$_2$=1.3) in a medium which is weakly damped ($\gamma_{0}=0.01$). In this study, the faster projectile particle (moving with M$_1$=1.3) is initially ($t=0$) placed at $y/\lambda_d=2.5$ whereas the slower projectile particle (moving with M$_1$=1.1) is placed at $y/\lambda_d=-2.5$ at a distance of $x/\lambda_d=10$, so that the trailing faster projectile particle can overtake the leading slower cone during the time window of interest. It is clear from the figure that the wake which is excited due the motion of the faster projectile particle trails the slower wake till the time $t=49\tau_d$ (see Fig.~\ref{figure5}(g) and (h)). Afterwards the trailing cone structure overtakes the leading cone structure. In the case of co-propagating Mach cones, the wings always interact with each other from the beginning of their journey but the interaction is only limited in the common region i.e, in the region between the two source points. As a result the formation of maxima and minima are seen in this common region unlike the earlier case when two Mach cones propagate in opposite directions. It is also seen that the interference patterns are brighter near the source point whereas they becomes weaker when one goes away from the two source points of the Mach cone. In the case of two co-propagating Mach cones, fewer constructive/destructive interference patterns are observed in $y/\lambda_d$ direction, whereas in the case of counter-propagating  Mach cone (see Fig.~\ref{figure1}) one can obtain more (maximum up to five) fringes. 
\section{Conclusion}
\label{sec:conclusion}
In conclusion, we have theoretically investigated the interaction characteristics  of two counter/co propagating Mach cones generated by charged projectile particles moving with supersonic velocities in opposite/same directions in a dusty plasma medium. Analytic expressions for the perturbed densities and potentials of the Mach structures have been obtained from a solution of a set of linearized model fluid equations for the dust dynamics. The contributions of the charged projectiles have been incorporated in the Poisson equation.  The time evolution of the Mach cones and their mutual interactions have been studied by numerical evaluations of the analytic expressions for varying conditions of the projectile velocities and directions of travel. The interacting Mach cones are seen to create interesting and long lasting interference patterns in density perturbations whose characteristics are determined by the relative velocities of the projectiles as well as by the medium properties. In particular the inter-fringe distance  of the pattern is seen to increase monotonically with the relative velocity whereas the inclination of the fringe is found to decrease with the relative velocity.  Measurement of such signatures in patterns between interacting charged projectiles can serve as a useful diagnostic tool for estimating the relative velocity of the projectiles. A potential application of such a tool could be in astrophysical scenarios like planetary rings where interactions between Mach cones excited by fast moving large sized charged objects moving in an environment of finer charged dust particles and electron ion plasma are likely to occur. An experimental verification of our results would be useful in developing such a tool. It can be easily carried out in present day laboratory set-ups by re-doing some of the past Mach cone experiments in dusty plasmas \cite{Samsonov1,Samsonov2,Jiang2,Mierk} in the presence of two counter/co travelling fast particles. Such a verification can provide an appropriate benchmark for a diagnostic tool that can then be applied to other situations. The experiment could also serve to delineate  deviations if any arising from possible nonlinear features of Mach cones - a physics that has not been included in our present analysis and that remains an open problem for future investigations. \\


\end{document}